\documentclass[aps,prl,twocolumn,superscriptaddress,groupedaddress]{revtex4-2}
\usepackage{amsmath,amssymb,amsfonts}
\usepackage{bm, latexsym}
\usepackage[dvipdfmx]{graphicx}
\usepackage{bbm,times}
\usepackage[normalem]{ulem}
\usepackage[usenames,dvipsnames]{color}
\usepackage[pdftex]{hyperref}
\usepackage{braket}

\usepackage{comment}

\newcommand{\cred}[1]{\textcolor{black}{#1}}

\graphicspath{
{../image/}
}

\begin{document}

\title{Theory of non-Hermitian fermionic superfluidity on a honeycomb lattice: Interplay between exceptional manifolds and van Hove singularity}
\author{Soma Takemori}
\email{s-takemori@stat.phys.titech.ac.jp}
\author{Kazuki Yamamoto}
\author{Akihisa Koga}
\affiliation{Department of Physics, Tokyo Institute of Technology,
  Meguro, Tokyo 152-8551, Japan}


\date{\today}

\begin{abstract}
  We study the non-Hermitian fermionic superfluidity subject to dissipation of Cooper pairs on a honeycomb lattice, for which we analyze the attractive Hubbard model with a complex-valued interaction. Remarkably, we demonstrate the emergence of the dissipation-induced superfluid phase that is anomalously enlarged by a cusp on the phase boundary. We find that this unconventional phase transition originates from the interplay between exceptional lines and van Hove singularity, which has no counterpart in equilibrium. Moreover, we demonstrate that the infinitesimal dissipation induces the nontrivial superfluid solution at the critical point.
  \cred{Our results can be tested in ultracold atoms with photoassociation techniques by postselcting special measurement outcomes with the use of quantum-gas microscopy and can lead to understanding the NH many-body physics triggered by exceptional manifolds in open quantum systems.}
\end{abstract}

\maketitle

\textit{Introduction}.---
Recent advancement in the experimental techniques of ultracold atoms has realized a feasible control of physical parameters and associated quantum state preparations in many-body physics \cite{schafer20}. As a practical tool for quantum simulation, rich phenomena in condensed matter physics have been successfully conducted by preparing an optical lattice, such as fermionic superfluidity \cite{chin06}, Mott insulators \cite{jordens08}, and nonequilibrium dynamics \cite{scott19,muniz20}. These experimental setups allow us to reach quantum many-body phases that have been inaccessible in solid-state systems. Moreover, ultracold atoms in an optical lattice have broadened their research area to the preparation of unconventional lattice structures, such as honeycomb \cite{tarruell12,jotzu14}, triangular \cite{ozawa23}, and kagome lattices \cite{jo12}. In particular, fermionic superfluidity, which is one of the most striking many-body physics, has been studied on such complex optical lattices, e.g., the BCS-BEC crossover \cite{zhao06}, the Fulde-Ferrell-Larkin-Ovchinnkikov superfluidity \cite{cichy18}, and spontaneous superflow induced by antiferromagnetic frustrations \cite{yang10}. These studies have opened a new field to investigate many-body physics in controllable setups.

On the other hand, as any physical system is surrounded by environments, it is of particular importance to identify how dissipation affects the quantum coherence of many-body states \cite{daley14, sieberer16}. Recent experimental progress in ultracold atoms has realized unprecedented many-body phenomena in open quantum systems \cite{syassen08,mark12,yan13,barontini13,zhu14,patil15,luschen17,tomita17,sponselee18,tomita19,corman19,takasu20,bouganne20,huang23,honda23}. Motivated by these experiments, much attention has been drawn to the non-Hermitian (NH) quantum many-body systems \cite{ashida20}, e.g., NH quantum phase transitions \cite{yamamoto19,hanai19,hanai20,hamazaki19,nakagawa20,matsumoto20,nakagawa21, yang21,zhang21eta, wang23,yu23,sarkar23,resendiz20,liu20,xu20,zhang20,sayyad23topo,sayyad23trans}, quantum critical phenomena \cite{ashida16,ashida17,nakagawa18,lourencco18,yamamoto22,yamamoto23sun,han23}, and measurement-induced entanglement dynamics under continuous monitoring \cite{fuji20,goto20,tang20,jian21SYK,block22,minato22,doggen22,yamamoto23local}. Notably, NH BCS theory has been proposed in Ref.~\cite{yamamoto19}, and many theoretical investigations have been performed so far to explore unconventional NH fermionic superfluid phase transitions associated with exceptional manifolds \cite{gahtak18,iskin21,kanazawa21,he21,li23,tajima23,mazza23}, which are the anomalous singularity unique to NH systems.
While the growing importance of studying how exceptional manifolds change conventional condensed matter physics, a unified understanding of their impact on NH quantum many-body phenomena still remains a challenging issue.

\begin{figure}[b]
  \centering
  \includegraphics[width = 8.5cm]{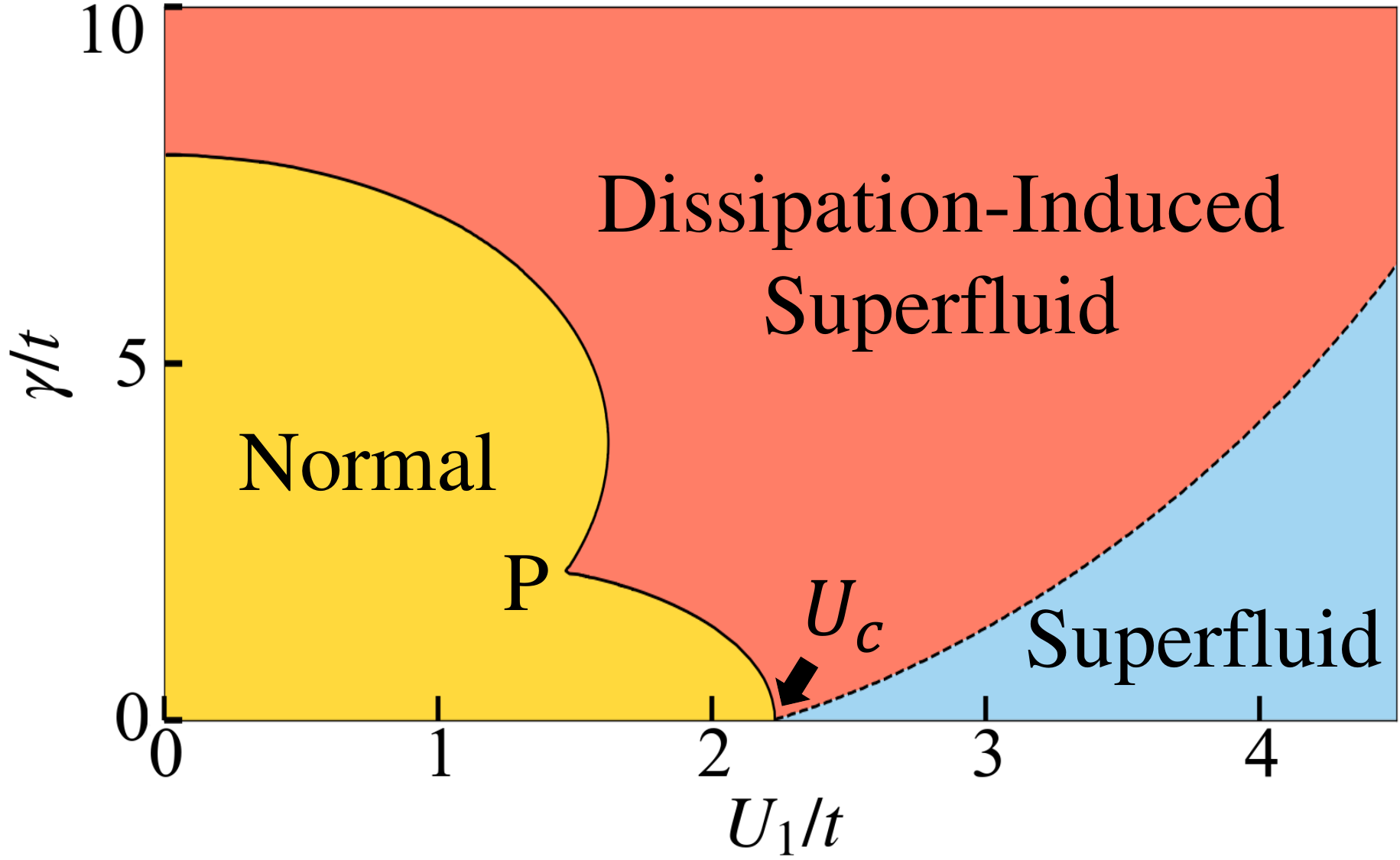}
  \caption{
    Phase diagram of the attractive Hubbard model with a complex-valued interaction.
    The cusp (P) with the double-humped behavior of the phase boundary between the normal and DS states originates from the interplay between ELs and VHS (see text).
  }
  \label{NHhoney_phasediagram_NHBCS_mu0_image}
\end{figure}

\cred{In this Letter, we tackle this problem by analyzing the NH fermionic superfluidity on a honeycomb lattice and elucidate the emergence of an unconventional NH many-body phase diagram, where the dissipation-induced superfluid (DS) phase is anomalously enlarged with a cusp on the phase boundary (see Fig.~\ref{NHhoney_phasediagram_NHBCS_mu0_image}), which has not been reported in the phase diagrams in the conventional equilibrium systems.}
To clarify the origin of the unconventional phase boundary, we examine the dominant contribution to the NH gap equation and find that the NH phase transition occurs when the exceptional lines (ELs) appear in the Brillouin zone. Because of this unique origin, the dissipative superfluid is fairly enhanced when ELs pass through van Hove singularity (VHS), resulting in the cusp with the double-humped behavior. This enhancement on the DS state has no counterpart in Hermitian system.

Moreover, we show that the infinitesimal dissipation yields
the nontrivial superfluid solution
when the system without dissipation is located at the phase boundary between the normal
and superfluid states.
Such mechanism is crucial in realizing the NH fermionic superfluidity
in the system with VHS
and reveals that exceptional manifolds can dominate the unconventional NH phase transitions and exotic quantum phenomena in open quantum systems. Our model can be experimentally realized in ultracold atoms by introducing two-body loss with photoassociation techniques and postselecting null measurement outcomes with quantum gas microscopy.

\begin{figure}[tb]
  \centering
  \includegraphics[width = 8.5cm]{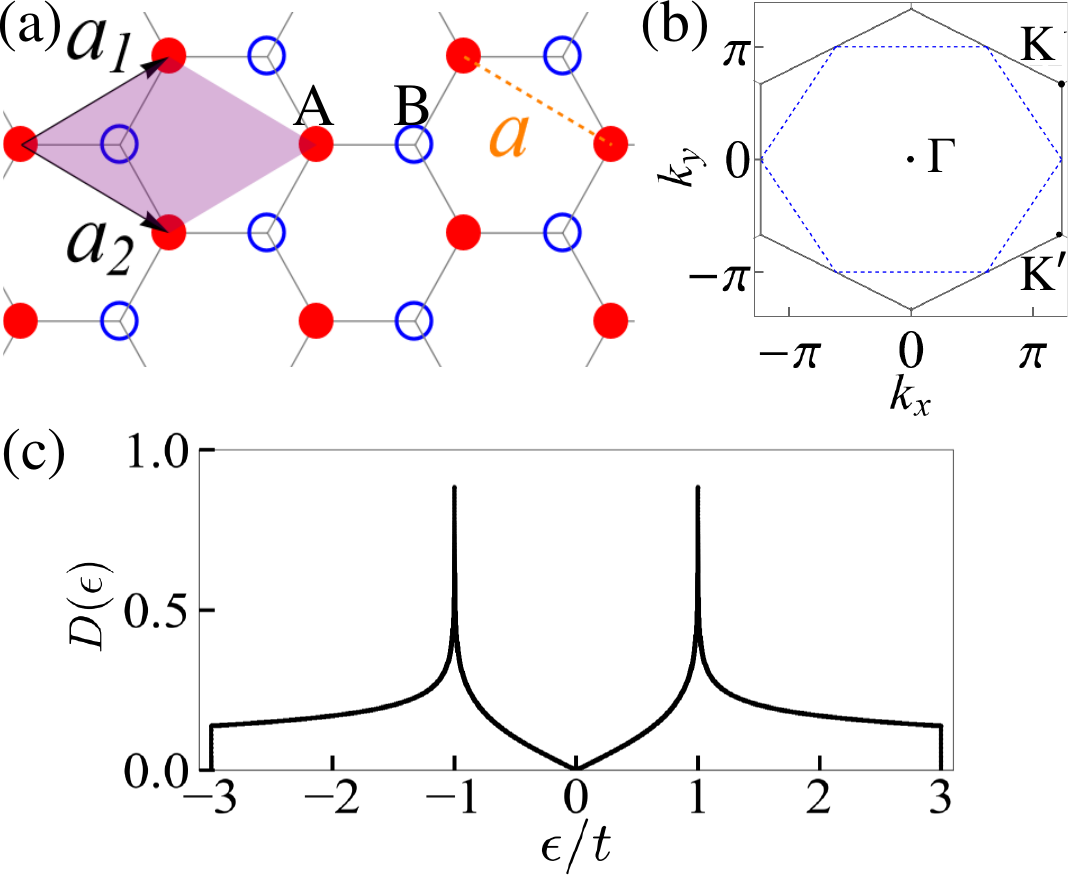}
  \caption{(a) Honeycomb lattice with two sublattices A and B,
    and primitive lattice vectors $\bm{a}_{1},\bm{a}_{2}$.
    (b) Corresponding first Brillouin zone.
    The Fermi points for the half-filled noninteracting system appear at K and K' points.
    (c) DOS for the noninteracting system. There exists a logarithmic VHS
    at $\epsilon/t = \pm 1$ and the corresponding line in Brillouin zone is
    shown as the dashed line in (b).}
  \label{NHhoneyEP_honeycomb_lattice_real_space_DOS_image}
\end{figure}

\textit{Model}.---We consider the effect of the two-body loss in ultracold fermions
with attractive interactions on a honeycomb lattice,
which is shown in Fig.~\ref{NHhoneyEP_honeycomb_lattice_real_space_DOS_image}(a).
The dissipative dynamics is governed by the Lindblad equation since a large internal energy is converted to kinetic energy and scattered atoms are immediately lost into surrounding environments \cite{daley14}. Then, we employ the quantum-trajectory method and postselect special outcomes of null measurement process with the use of quantum-gas microscopy \cite{ott16,mitra18,brown20,chan20,hartke23,ozawa23}. As a result, the system is described by the NH Hubbard Hamiltonian with a complex-valued interaction as
\begin{align}
   & H_{\text{eff}} = H_{0} - U\sum_{i,s}n_{i,s,\uparrow}n_{i,s,\downarrow}, \label{NHhoneyEP_effective_Hubbard_hamiltonian_irep_eq} \\
   & H_{0}          = \sum_{\bm{k},\sigma}\left(\epsilon_{\bm{k}}
  c_{\bm{k},A,\sigma}^{\dagger}c_{\bm{k},B,\sigma} + \text{H.c.}\right)
  - \mu \sum_{\bm{k},s,\sigma}n_{\bm{k},s,\sigma},
\end{align}
where $c_{i(\bm{k}),s,\sigma}$ annihilates a fermion with spin $\sigma = \uparrow,\downarrow$
at site $i$
[with momentum $\bm{k}$ in the first Brillouin zone, see Fig.~\ref{NHhoneyEP_honeycomb_lattice_real_space_DOS_image}(b)]
in the sublattice $s=A, B$, $c_{i(\bm{k}),s,\sigma}^\dag$ is
the corresponding creation operator,
and $n_{i(\bm{k}),s,\sigma}=c_{i(\bm{k}),s,\sigma}^{\dagger}c_{i(\bm{k}),s,\sigma}$  is the particle number operator.
The Fourier transformation of the operator is defined as
$c_{i,s,\sigma} = \sqrt{2/N}\sum_{\bm{k}}e^{i\bm{k}\cdot\bm{r}_{s}}c_{\bm{k},s,\sigma}$,
where $\bm{r}_{s}$ denotes the real-space position of the sublattice $s$,
and $N$ is the number of lattice sites.
The complex-valued interaction is introduced as $U=U_{1}+i\gamma/2$, where $U_{1}>0$ is a bare attractive interaction of the system and $\gamma>0$ denotes the strength of dissipation \cred{[For the detailed derivation of the NH Hamiltonian~\eqref{NHhoneyEP_effective_Hubbard_hamiltonian_irep_eq} from the Lindblad equation, see the Supplemental Material]}.
Here, $\epsilon_{\bm{k}}=-t(1+e^{-i\bm{k}\cdot\bm{a}_{1}}+e^{-i\bm{k}\cdot\bm{a}_{2}})$,
$\mu$ is the chemical potential of the system, and $t$ is the hopping amplitude .
The primitive lattice vectors on a honeycomb lattice $\bm{a}_{1} = a(\sqrt{3}/2,1/2)$ and $\bm{a}_{2} = a(\sqrt{3}/2,-1/2)$ are illustrated
in Fig.~\ref{NHhoneyEP_honeycomb_lattice_real_space_DOS_image}(a).
Hereafter, we take $t$ as the energy unit.

In the noninteracting case $U=0$, the dispersion relations are given
by $\pm |\epsilon_{\bm k}|$, and the corresponding density of states (DOS) is shown in Fig.~\ref{NHhoneyEP_honeycomb_lattice_real_space_DOS_image}(c).
We note that the noninteracting system has two important features.
One of them is that the linear dispersion appears around K and K' points,
and the DOS is absent at $\epsilon=0$.
Therefore, the finite critical interaction $U_{c}$ is necessary to drive the half-filled system
to the superfluid states~\cite{sorella92,furukawa01,assaad13},
which is in contrast to the system on the square lattice and the cubic lattice with $U_{c}=0$.
The other is the existence of VHS at $\epsilon=\pm 1$,
which is clearly found in Fig.~\ref{NHhoneyEP_honeycomb_lattice_real_space_DOS_image}(c).
In the Hermitian system, this little affects the superfluid phase. However, VHS plays a crucial role in the NH system, which will be discussed later.

\textit{Non-Hermitian BCS theory on a honeycomb lattice}.---
To describe the superfluid state in the framework of the NH mean-field theory~\cite{yamamoto19},
we introduce the superfluid order parameters as
\begin{align}
  \Delta       & = -\frac{2U}{N}\sum_{\bm{k}}{}_{L}\langle c_{-\bm{k},s,\downarrow}c_{\bm{k},s,\uparrow}\rangle_{R}, \label{NHhoneyEP_delta_def_eq}                        \\
  \bar{\Delta} & = -\frac{2U}{N}\sum_{\bm{k}}{}_{L}\langle c_{\bm{k},s,\uparrow}^{\dagger}c_{-\bm{k},s,\downarrow}^{\dagger}\rangle_{R}, \label{NHhoneyEP_bardelta_def_eq}
\end{align}
where ${}_{L}\langle O\rangle_{R} \equiv {}_{L} \langle \text{BCS}|O|\text{BCS}\rangle_{R}$ is the expectation value of the operator $O$, and $| \text{BCS}\rangle_{L(R)}$ is
the left (right) BCS state.
Here, we have assumed that the superfluid order parameters are independent of the sublattices.
\cred{We also note that the order parameter $\Delta_0$ is not directly related to an energy gap in an NH system~\cite{gong18}.}
Then, the NH BCS Hamiltonian is written as
\begin{align}
  H_{\text{eff}}^{\text{BCS}} & = H_{0} - \sum_{\bm{k},s} [\Delta c_{\bm{k},s,\uparrow}^{\dagger}c_{-\bm{k},s,\downarrow}^{\dagger} + \bar{\Delta} c_{-\bm{k},s,\downarrow}c_{\bm{k},s,\uparrow} ] + \frac{N\Delta\bar{\Delta}}{U} \\
                              & =\sum_{\bm{k},\alpha,\sigma} E_{\bm{k},\alpha}
  \left(
  \bar{f}_{\bm{k},\alpha,\uparrow}f_{\bm{k},\alpha,\uparrow}+
  \bar{f}_{-\bm{k},\alpha,\downarrow}f_{-\bm{k},\alpha,\downarrow}\right)+E_{0},  \label{NHhoneyEP_NH_BCS_hamiltonian_eq}
\end{align}
where $f_{\bm{k},\alpha,\sigma}$ ($\bar{f}_{\bm{k},\alpha,\sigma}$) is an annihilation (creation) operator for the quasiparticle, the subscripts $\alpha=\pm$ are defined for the dispersion relation $E_{\bm{k},\pm} = \sqrt{(|\epsilon_k|\pm \mu)^{2}+\Delta\bar{\Delta}}$, and $E_{0} = N\Delta \bar{\Delta}/U - \sum_{\bm{k},\alpha}(E_{\bm{k},\alpha}+\mu)$ is the energy of the NH BCS state. Then, the NH BCS states are defined by the vacuum states for the quasiparticle operators and are given by
\begin{align}
  |\text{BCS}\rangle_{R} & = \prod_{\bm{k},\alpha}\Big(u_{\bm{k},\alpha}+v_{\bm{k},\alpha} d_{\bm{k},-\alpha,\uparrow}^{\dagger}d_{-\bm{k},-\alpha,\downarrow}^{\dagger}\Big)|0\rangle, \label{NHhoneyEP_RightBCSgroundstate_def_eq}                     \\
  |\text{BCS}\rangle_{L} & = \prod_{\bm{k},\alpha}\Big(u_{\bm{k},-\alpha}^{\ast} + \bar{v}_{\bm{k},-\alpha}^{\ast}d_{\bm{k},\alpha,\uparrow}^{\dagger}d_{-\bm{k},\alpha,\downarrow}^{\dagger}\Big)|0\rangle, \label{NHhoneyEP_LeftBCSgroundstate_def_eq} 
\end{align}
where we have introduced $d_{\bm{k},\pm,\uparrow}^{\dagger}=(c_{\bm{k},A,\uparrow}^{\dagger} \pm |\epsilon_{\bm{k}}|/\epsilon_{\bm{k}} c_{\bm{k},B,\uparrow}^{\dagger})/\sqrt{2}$ and $d_{-\bm{k},\pm,\downarrow}=(c_{-\bm{k},A,\downarrow} \pm |\epsilon_{\bm{k}}|/\epsilon_{\bm{k}} c_{-\bm{k},B,\downarrow})/\sqrt{2}$.
Here, the coefficients satisfy $u_{\bm{k},\pm}^{2} + v_{\bm{k},\pm}\bar{v}_{\bm{k},\pm} = 1$, and
the state $|0\rangle$ is the vacuum for fermions \cite{Supple}.

Finally, by substituting Eqs.~\eqref{NHhoneyEP_RightBCSgroundstate_def_eq} and \eqref{NHhoneyEP_LeftBCSgroundstate_def_eq} into Eqs.~\eqref{NHhoneyEP_delta_def_eq} and \eqref{NHhoneyEP_bardelta_def_eq}, we obtain the NH gap equation for a nontrivial superfluid phase as
\begin{equation}
  \frac{N}{U} = \sum_{\bm{k},\alpha}\frac{1}{2E_{\bm{k},\alpha}}. \label{NHhoneyEP_gap_equation_eq}
\end{equation}
In the following, we fix the special U(1) gauge so that the relation $\Delta=\bar{\Delta}=\Delta_{0}\in \mathbb C$ is satisfied.
In this case, the order parameter is smoothly connected to the real value when one approaches the Hermitian limit \cite{yamamoto19}.
We also assume the half-filled condition $\mu=0$, where $E_{\bm{k},+}=E_{\bm{k},-}=E_{\bm{k}}$ and the expectation value of the particle number satisfies $\sum_{i,s,\sigma}{}_{L}\langle n_{i,s,\sigma}\rangle_{R}=N$ \cite{yamamoto19}.
\cred{The chemical potential is fixed since we focus on the conditional dynamics where the particle number is unchanged by postselecting the special measurement outcomes.}
\cred{We also note that our formalism of the BCS theory for a dissipative superfluid [Eqs.~\eqref{NHhoneyEP_delta_def_eq}-\eqref{NHhoneyEP_gap_equation_eq}] is equivalent to that of the saddle point approximation in the path-integral approach~\cite{yamamoto19}.}

\textit{Unconventional superfluid phase transitions}.---
By solving the gap equation \eqref{NHhoneyEP_gap_equation_eq} and
examining the condensation energy
$E_{\text{cond}}=N\Delta_{0}^{2}/U - \sum_{\bm{k},\alpha}(E_{\bm{k},\alpha} - |\epsilon_{\bm{k}}|)$
of the superfluid state,
we obtain the phase diagram as illustrated in Fig.~\ref{NHhoney_phasediagram_NHBCS_mu0_image}.
\cred{The normal state, where the gap equation only gives the trivial solution $\Delta_{0}=0$, is shown as the yellow region. Although the dissipation for weak attraction never induces the superfluid state, there is a nontrivial solution with finite $\Delta_{0}$ for large dissipation, which typically destroys the coherence of the quantum state. Then, we define the red region where $\Delta_{0}\neq 0$ and $\text{Re} E_{\text{cond}}>0$ as the DS state. In the blue region, the superfluid state emerges with finite $\Delta_{0}$ and $\text{Re} E_{\text{cond}}<0$. We emphasize that a nonzero real part of the superfluid gap means the emergence of the superfluid (DS) state in $\text{Re}E_{\text{cond}} < 0 (\text{Re}E_{\text{cond}} > 0)$.
We also remark that the imaginary parts of the superfluid gap and the condensation energy lead to the instability of the superfluid. In the experiment, these two phase boundaries can be detected by means of the continuous quantum Zeno effect (QZE) and Feshbach resonance.}


Examining the superfluid gap and condensation energy,
we clarify the nature of the superfluid state inherent in the NH system and
discuss the phase transitions among the above three states.
When the strength of dissipation is strong enough, the DS state is realized.
\cred{This is managed by the QZE~\cite{syassen08,mark12,barontini13,yan13,zhu14,tomita17}, which suppresses the hopping of the atoms to neighboring sites. Then, in contrast to the Hermitian case where the particles move freely, particles are bounded due to strong dissipation even in weak attraction and can form a pair.}

\begin{figure}[bt]
  \centering
  \includegraphics[width = 8.5cm]{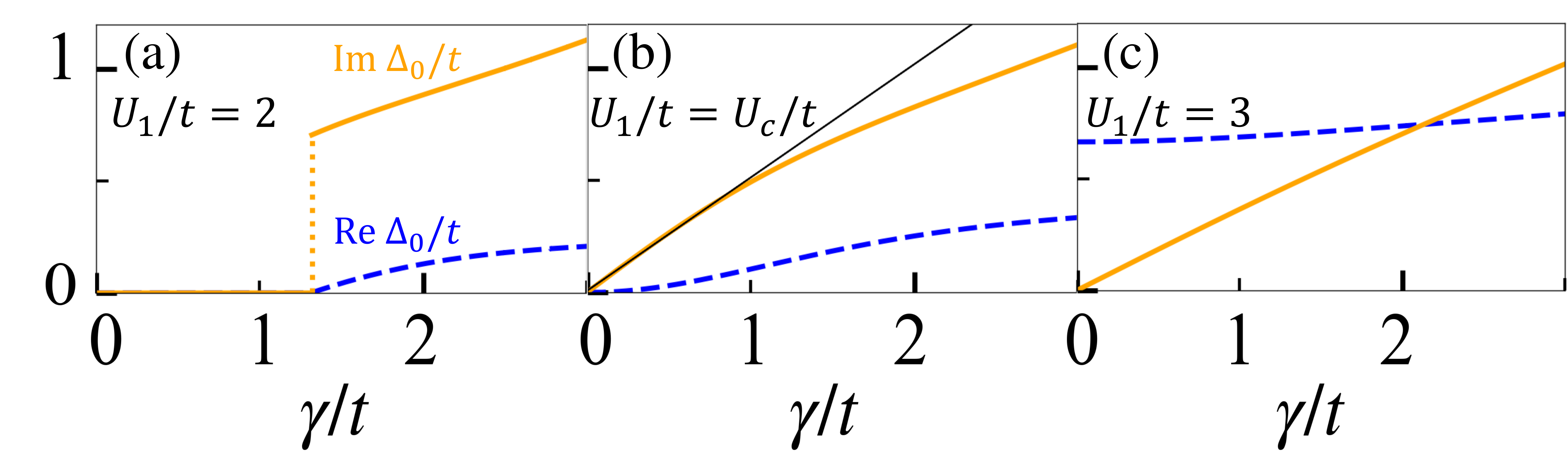}
  \caption{Dashed (solid) lines represent the real (imaginary) part of
  the order parameter $\Delta_0$
  when (a) $U_{1}/t=2$, (b) $U_{1}/t=U_{c}/t (=2.23)$, and (c) $U_1/t=3$. \cred{The imaginary part of the superfluid gap first increases linearly with increasing dissipation [black line in (b)].}
  }
  \label{NHhoney_EP_gap_U2.23_image}
\end{figure}

When $U_{1}/t=2$,
the decrease of the dissipation induces the unconventional phase transition from the DS state
to the normal state at $\gamma_{c}/t = 1.32$,
as shown in Fig.~\ref{NHhoney_EP_gap_U2.23_image}(a).
We emphasize that approaching the phase boundary from the strong dissipation regime,
the real part of the superfluid gap gradually decreases while the imaginary part remains finite.
Then, the superfluid gap completely disappear for $\gamma < \gamma_{c}$.
This means that the jump singularity appears in the imaginary part of $\Delta_0$
and the ELs appear in the Brillouin zone, which will be discussed in detail.
\cred{This unconventional phase transition associated with ELs is unique to the NH systems.}
When the system approaches $(U_{1},\gamma)=(U_c,0)$ along the phase boundary,
the jump in the imaginary part of $\Delta_{0}$ decreases.
To clarify the phase transition around the critical point,
we use the simple linear DOS
which can capture low energy properties inherent in the honeycomb lattice (e.g.
the Dirac dispersion relations).
The gap equation with the linear DOS demonstrates that
the phase boundary $\gamma \propto \sqrt{U_{c}-U_{1}}$.
This means that the NH gap equation has a non-trivial solution for $|U|> U_{c}$ \cite{Supple}.
Then, at the critical point,
the DS solution appears
under infinitesimal dissipation rate
and the imaginary part of $\Delta_0\propto \gamma$.
This is consistent with the numerical result
as shown in Fig.~\ref{NHhoney_EP_gap_U2.23_image}(b). This indicates that the characteristic phenomena around the critical point are described by the Dirac dispersion relations even in the NH system \cite{Supple}. We emphasize that this transition is distinct from that of the cubic lattice \cite{yamamoto19}, where a finite strength of dissipation is necessary to cause a phase transition between the normal and the DS states.
\cred{We note that $\Delta_{0}=U/2$ in strong dissipation limit, which means that the atoms form the pair on the same site due to QZE. This is similar to the strong-coupling BEC-like superfluid state in Hermitian system.}


When $U_{c}< U_{1}$, distinct behavior appears in the order parameter.
Figure~\ref{NHhoney_EP_gap_U2.23_image}(c) shows that increasing the dissipation,
the real part of $\Delta_0$ slowly increases and
the imaginary part of that is linearly induced.
We also find this nontrivial solution with finite $\Delta_0$ is robust for any strength
of dissipation.
This is because fermion pairs are tightly coupled on the same site
for strong attraction irrespective of the dissipation strength~\cite{yamamoto19}.
Further increase of dissipation induces the phase transition from the superfluid state
to the DS state
(see Fig.~\ref{NHhoney_phasediagram_NHBCS_mu0_image}).

\textit{Origin of the exotic phase diagram: Interplay between ELs and VHS}.---
Exceptional manifolds are the singularities
where not only eigenvalues but also eigenstates of the effective Hamiltonian are degenerate, and known to occur in NH fermionic superfluids \cite{yamamoto19,gahtak18,iskin21,kanazawa21,he21,li23,tajima23,mazza23}.
However, as shown in Fig.~\ref{NHhoney_phasediagram_NHBCS_mu0_image},
the phase boundary between the normal and DS states
displays a cusp associated with double-humped behavior, which has never been reported in the previous studies.
To clarify this anomalous extension of the DS region in the phase diagram,
we analyze how this phase boundary is affected by ELs inherent in the NH system.
When $\text{Re} \Delta_{0}/t=0$ at the transition point, the quasiparticle energy $E_{\bm{k}}$ becomes zero for $\text{Im} \Delta_{0}=\pm|\epsilon_{\bm{k}}|$ and ELs emerge. Notably, the momentum $\bm{k}_{0}$ that satisfies $E_{\bm{k}_{0}}=0$ significantly contributes to the gap equation
because the term $1/E_{\bm{k}_{0}}$ diverges.
An important point is that not only ELs but also noninteracting DOS leads to a large contribution to the right hand of the gap equation $\sum_{\bm{k}} 1/E_{\bm{k}}$.
In the following, we point out that this fact profoundly affects
the superfluid gap via the interplay between ELs and DOS
as ELs overlap the line for the VHS with changing dissipation.
We recall that the DOS away from the K and K' points plays a minor role in stabilizing the superfluid state in the Hermitian limit.

\begin{figure}[bt]
  \centering
  \includegraphics[width = 8.5cm]{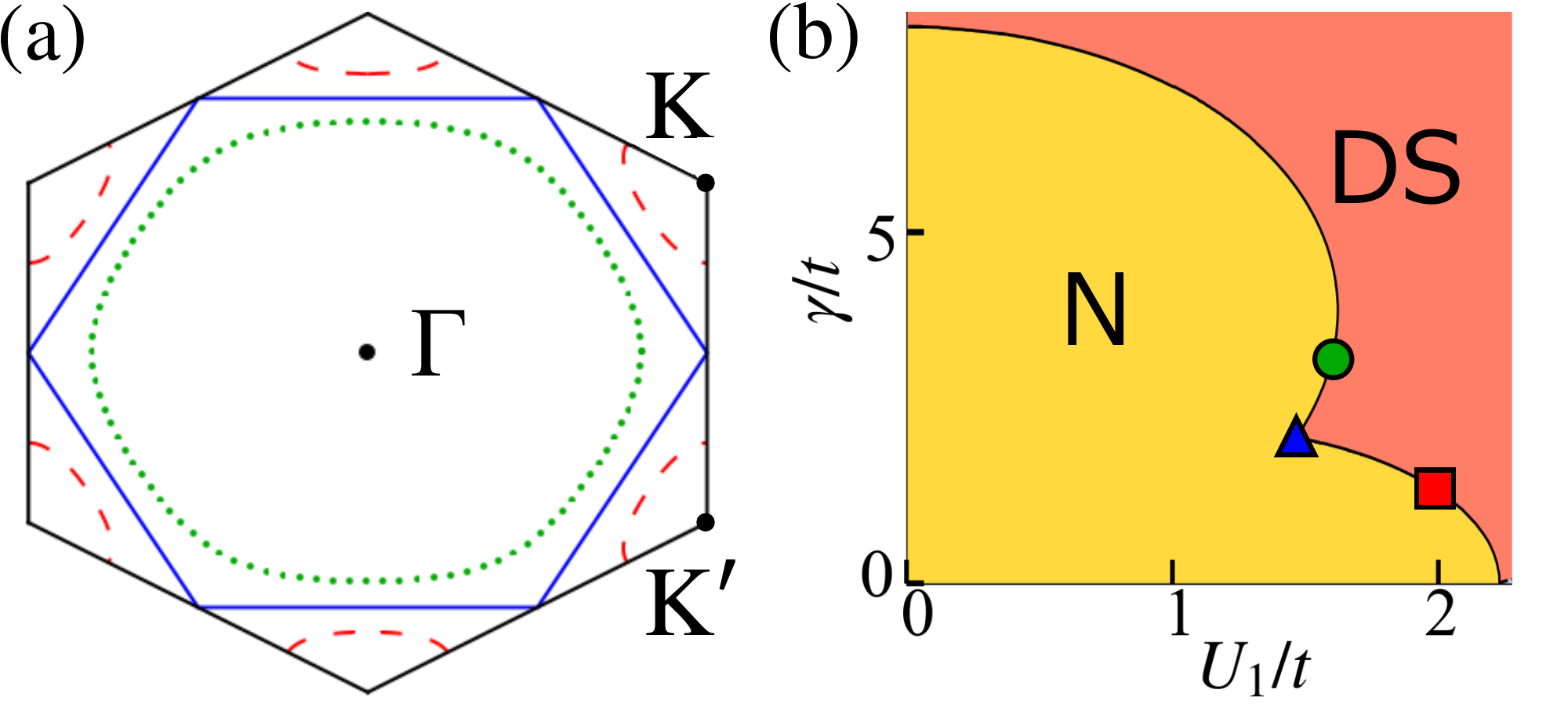}
  \caption{
    (a) Dashed, solid, and dotted lines present the ELs
    when $\Delta_{0}/t=0.7i$, $i$, and $1.3i$, respectively.
    The corresponding phase transition points are explicitly shown as the square, triangle, and circle in the phase diagram (b), respectively.
  }
  \label{NHhoney_EP_quasiE_Brillouin_image}
\end{figure}

To clarify how ELs and DOS impact on the phase diagram,
we illustrate ELs in momentum space
in Fig.~\ref{NHhoney_EP_quasiE_Brillouin_image}(a).
When the system is located at the critical point $(U_1, \gamma)=(U_c,0)$,
the ELs are located at K and K' points.
As increasing dissipation along the phase boundary,
ELs appear around K and K' points,
which are shown as the dashed lines in Fig.~\ref{NHhoney_EP_quasiE_Brillouin_image}(a).
In this case, the momentum close to the K and K' points plays an important role.
As already mentioned,
this means that
the phase transition between the normal and DS states
is gorvened by the linear DOS as is reminiscent of the Hermitian case.
Remarkably, for intermediate strength of dissipation with $\Delta_{0}/t=i$,
ELs completely match the line for the VHS at $|\epsilon_{\bm{k}}|/t=1$,
which is shown as the solid line in Fig.~\ref{NHhoney_EP_quasiE_Brillouin_image}(a).
This causes an anomalous cusp on the phase boundary,
which leads to exotic superfluid phase transitions such as the reentrant superfluidity
induced by dissipation.
\cred{We can say that this exotic phase diagram originates from the VHS and ELs inherent in the NH system on the honeycomb lattice.}
Moreover, we see that further increase of dissipation leads to
ELs inside of the Brillouin zone, which are shown as the dotted line
in Fig.~\ref{NHhoney_EP_quasiE_Brillouin_image}(a).
This means that the momentum far away from the K and K' points dominates the physics of the NH phase transition, resulting in the unconventional DS induced by QZE.
As a result, such an interplay between ELs and VHS brings about profound contributions to the NH gap equation \eqref{NHhoneyEP_gap_equation_eq} and leads to the unconventional phase transition characterized by the cusp on the phase boundary shown in Fig.~\ref{NHhoney_phasediagram_NHBCS_mu0_image}. This is nothing but the unique NH phase transition induced by the interplay between exceptional manifolds and many-body NH physics.
The characteristic shape of this phase boundary, which can be a smoking gun for the NH phase transition associated with exceptional manifolds, can be utilized to observe the NH fermionic superfluidity on a honeycomb optical lattice.

\textit{Towards experimental realization.}---
In order to observe the unconventional phase transition associated with a cusp of DS phase, we have to introduce two-body losses to fermions on a honeycomb optical lattice. \cred{On the other hand, as DS typically requires large dissipation, this seems to invalidate the no-jump dynamics that realizes NH superfluids.} However, since the physics behind DS for strong dissipation is QZE, the signature of the unconventional NH phase transition can be observed by using the following protocol~\cite{Supple}.

As the timescale of the relaxation towards superfluid states is governed by the inverse rate of the hopping of atoms $1/t$, we can observe the NH superfluid for the dissipation rate up to $\gamma\simeq t$ in order to postselect no-jump dynamics. In our model, the quantum phase transition from the normal state to the DS state occurs up to the dissipation rate $\gamma /t \simeq \mathrm{O}(1)$. \cred{By using this fact, we first prepare a stable superfluid for large attraction and introduce strong dissipation with the use of photoassociation techniques~\cite{tomita17, honda23}, so that the system goes into deep inside the DS phase.} \cred{Then, we decrease the interaction strength and the dissipation rate to observe DS around the phase boundary near the cusp~\cite{Supple}. As atom pairs are confined to each lattice site because of QZE, the dissociation of molecules can be delayed, and this fact can be utilized for the signature of the emergence of DS around the cusp.} \cred{This means that the NH superfluid transition associated with a cusp induced by ELs and VHS can be observed in experiments.}

As a current experimental situation, fermionic superfluidity has been realized in $^{40}\mathrm{K}$ \cite{regal04,stewart08} and $^{6}\mathrm{Li}$ atoms \cite{zwierlein04,bartenstein04,chin04} and the interaction strength can be tuned with the magnetic field for a Feshbach resonance, which can be operated fast on the order of $100\mu$s on an optical lattice \cite{chin06}. In order to obtain the NH dynamics, we have to postselect special measurement outcomes where the quantum jump process does not occur, and such dynamics is employed with the quantum-gas microscopy \cite{ozawa23,ott16,mitra18,brown20,chan20,hartke23}. Therefore, we believe that NH BCS superfluidity on a honeycomb lattice can be realized with experiments, which have also achieved unconventional lattice structures such as honeycomb, triangular, and kagome lattices \cite{tarruell12,jotzu14,ozawa23,jo12}.

\textit{Conclusion.}---
We have investigated the NH fermionic superfluidity on a honeycomb lattice.
Remarkably, we have shown that the nonmonotonic behavior of the phase boundary associated with a cusp between the normal and DS states emerges,
which originates from the interplay between ELs and VHS.
We emphasize that this characteristic behavior 
is ubiquitous in open quantum systems having VHS and can be generalized to other NH many-body phase transitions associated with exceptional manifolds (for the NH phase diagram on a square lattice, see the Supplemental Material). \cred{We expect that in other systems with VHS, similar phase diagrams with cusp are obtained.}
As we have focused on the mean-field analysis, it is of importance to employ more sophisticated analysis like exact methods, such as the Bethe ansatz in one dimension \cite{nakagawa21,sayyad23topo,sayyad23trans} and the Monte Carlo analysis \cite{yu23}.
\cred{Furthermore, the BCS theory is generalized to incorporate the change of the particle number described by quantum jump terms in the Lindblad equation \cite{yamamoto19}. It is interesting to explore the difference between the phenomena described by the NH BCS theory in our model and those described by the dissipative BCS theory following the Lindblad equation \cite{yamamoto21,dai23,xiao23},  which is left for furture work.}
\cred{Exploring the role of the higher order VHS with power law divergence~\cite{yuan19} in NH system may be another interesting work.}
We believe that this paper stimulates further study on exploring more abundant many-body phase diagrams induced by exceptional manifolds in open quantum systems.

\begin{acknowledgments}
  We are grateful to Masaya Nakagawa and Norio Kawakami for fruitful discussions.
  This work was supported by Grant-in-Aid for Scientific Research
  from JSPS, KAKENHI Grants No. JP23K19031 (K.Y.), No. JP19H05821, No. JP21H01025 and No. JP22K03525 (A.K.).
  K.Y. was also supported by Yamaguchi Educational and Scholarship Foundation.
\end{acknowledgments}

\nocite{apsrev42Control}
\bibliographystyle{apsrev4-2}
\bibliography{NHhoneyEP}


\clearpage

\renewcommand{\thesection}{S\arabic{section}}
\renewcommand{\theequation}{S\arabic{equation}}
\setcounter{equation}{0}
\renewcommand{\thefigure}{S\arabic{figure}}
\setcounter{figure}{0}

\onecolumngrid
\appendix
\begin{center}
  \large{Supplemental Material for}\\
  \textbf{``Theory of Non-Hermitian Fermionic Superfluidity on a Honeycomb Lattice: Interplay between Exceptional Manifolds and Van Hove Singularity"}
\end{center}

\section{\cred{The derivations of the non-Hermitian Hamiltonian}}

\cred{We consider the ultracold fermionic atoms with an attractive interaction in an optical lattice. Assuming that the atoms undergo inelastic collisions, the scattered atoms are quickly lost from the optical lattice because a large internal energy is converted into the kinetic energy. The full dissipative dynamics that can change the particle number of the system is described by the following Markovian Lindblad equation \cite{daley14}}
\begin{equation}
  \cred{\dot{\rho} = -i[H,\rho] - \frac{\gamma}{2}\sum_{i}\sum_{i}(L_{i}^{\dagger}L_{i}\rho + \rho L_{i}^{\dagger}L_{i} - 2L_{i}\rho L_{i}^{\dagger}),}\label{NHhoneySupple_Lindblad_eq}
\end{equation}
\cred{where $\rho$ is the density matrix and $H$ is the Hermitian Hamiltonian of the system. The Lindblad operator $L_{i}=c_{i\downarrow}c_{i\uparrow}$ describes two-body loss at site $i$ with rate $\gamma$. As Markovian master equation typically requires that the system is weakly coupled to the environment, the relaxation timescale of the environment should be faster than that of the system. This requires that the energy scale of the environment, which is determined by the energy scale of particles escaping from the system due to inelastic collisions, should be larger than that of the system.}
\cred{The Lindblad equation~\eqref{NHhoneySupple_Lindblad_eq} is rewritten as}
\begin{equation}
  \cred{\dot{\rho} = -i(H_{\text{eff}}\rho - \rho H_{\text{eff}}) + \gamma\sum_{i}L_{i}\rho L_{i}^{\dagger},} \label{NHhoneySupple_lindblad_Heff_and_quantumjump_description_eq}
\end{equation}
\cred{where the non-Hermitian Hamiltonian is given by $H_{\text{eff}} = H - i\gamma/2\sum_{i}L_{i}^{\dagger}L_{i}$. Considering the standard Hubbard model $H=\sum_{\bm{k},\sigma}(\epsilon_{\bm{k}}c_{\bm{k},A,\sigma}^{\dagger}c_{\bm{k},B,\sigma} + \text{H.c.}) -\mu\sum_{\bm{k},s,\sigma}n_{\bm{k},s,\sigma} - U_{1}\sum_{i,s}n_{i,s,\uparrow}n_{i,s,\downarrow}$ and the two-body loss denoted as $L_{i}=c_{i,\downarrow}c_{i,\uparrow}$, the non-Hermitian Hamiltonian is rewritten as}
\begin{equation}
  \cred{H_{\text{eff}} = H - \frac{i\gamma}{2}\sum_{i}c_{i,\uparrow}^{\dagger}c_{i,\downarrow}^{\dagger}c_{i,\downarrow}c_{i,\uparrow} = \sum_{\bm{k},\sigma}(\epsilon_{\bm{k}}c_{\bm{k},A,\sigma}^{\dagger}c_{\bm{k},B,\sigma} + \text{H.c.}) -\mu\sum_{\bm{k},s,\sigma}n_{\bm{k},s,\sigma} - (U_{1}+\frac{i\gamma}{2}) \sum_{i,s}n_{i,s,\uparrow}n_{i,s,\downarrow}} \label{NHhoneySupple_nonHermitianHubbard_model_eq}
\end{equation}
\cred{The effect of the dissipation is taken into account in the NH Hamiltonian as a complex-valued interaction.}

\begin{figure}[h]
  \centering
  \includegraphics[width=12cm]{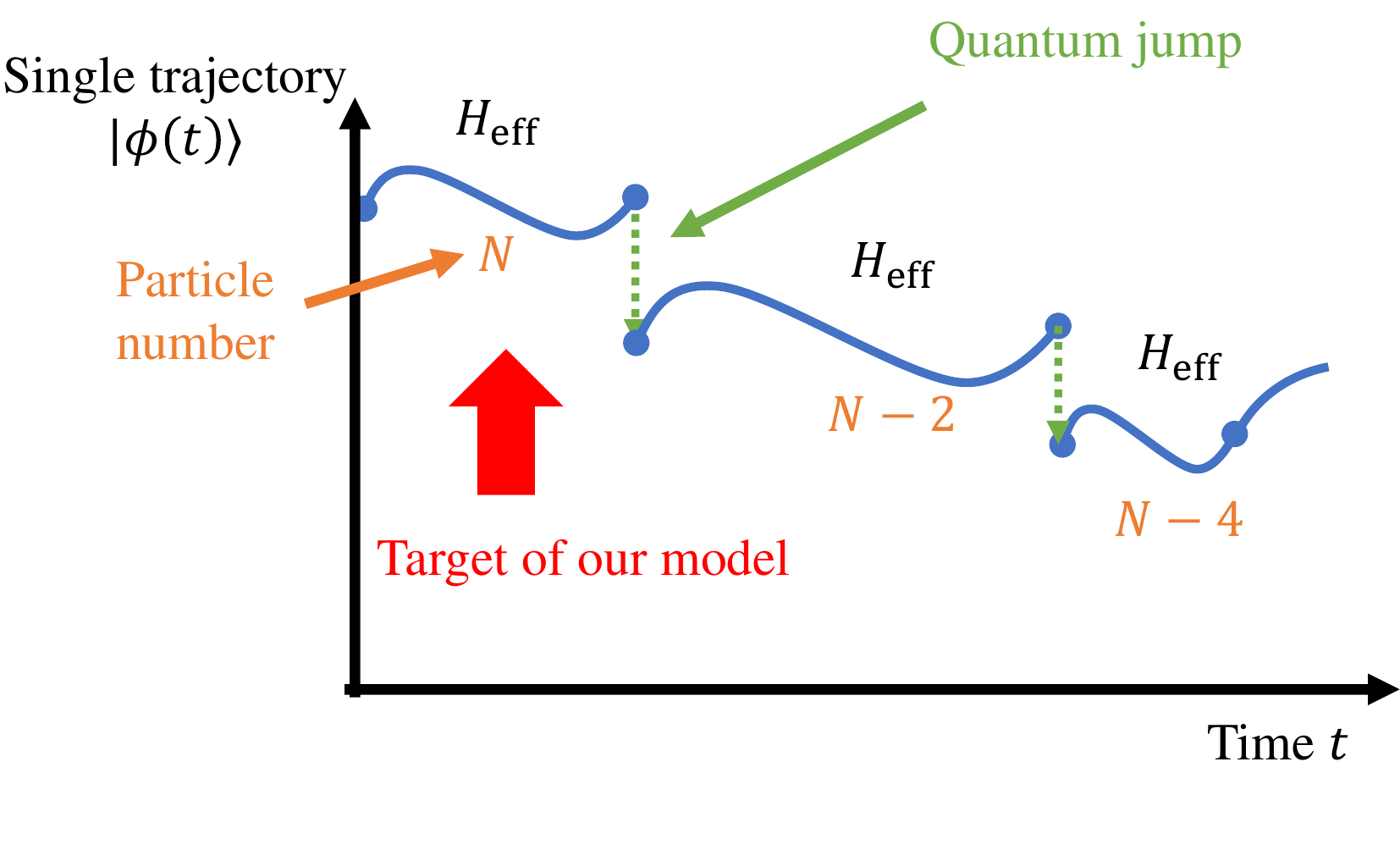}
  \caption{\cred{The schematic image of a single trajectory. Each trajectory involves nonunitary time evolution under $H_{\text{eff}}$ and the stochastic quantum-jump process described by $L_{i}=c_{i\downarrow}c_{i\uparrow}$. In our present paper, we focus on the dynamics of the null measurement process described by $H_{\text{eff}}$ indicated by a red arrow.}}
  \label{NHhoneySupple_quatnum_trajectory_image}
\end{figure}

\cred{To derive the effective Hamiltonian $H_{\text{eff}}$, we use the quantum trajectory method~\cite{daley14} where the quantum-trajectory is the time evolution obtained by unraveling the Lindblad equation~~\eqref{NHhoneySupple_Lindblad_eq}. The time evolution of each trajectory is schematically illustrated in Fig.~\ref{NHhoneySupple_quatnum_trajectory_image}.}

\cred{The single trajectory involves two parts of the stochastic propagation: nonunitary schr\"{o}dinger equation under the effective NH Hamiltonian $H_{\text{eff}}=H-i\gamma/2\sum_{i}L_{i}^{\dagger}L_{i}$ (Blue line in Fig.~\ref{NHhoneySupple_quatnum_trajectory_image}) and the stochastic quantum-jump process which describes the change in the particle number (green dotted line in Fig.~\ref{NHhoneySupple_quatnum_trajectory_image}). The last term of the right-hand side of Eq.~\eqref{NHhoneySupple_lindblad_Heff_and_quantumjump_description_eq} describes the quantum-jump process. When the system initially has the particle number $N_{0}$, the particle number will be changed to $N_{0}-2,N_{0}-4,\cdots$ since the quantum-jump sometimes occurs. Then, by postselecting the dynamics of the system which does not include the jump process (indicated by red arrow in Fig.~\ref{NHhoneySupple_quatnum_trajectory_image}), the time evolution is effectively described by $H_{\text{eff}}$ and the particle number of the system does not change. Assuming that the half-filling condition is achieved in the initial state, the chemical potential is fixed to $\mu=0$.}

\section{The detailed calculations of the NH BCS theory on a honeycomb lattice}


Here, we explain the detailed calculation of the NH BCS theory on a honeycomb lattice. By using the general relation for the order parameters given in Ref.~\cite{yamamoto19}, we can define $2\times 2$matrices $H_{11} = -H_{22} = -\mu I + (\text{Re} \epsilon_{\bm{k}})\sigma_{x} - (\text{Im} \epsilon_{\bm{k}}) \sigma_{y}, H_{12} = \Delta I, H_{21} = \bar{\Delta} I$ for the NH attractive Hubbard model [Eq. (1) in the main text], where $I$ is the identity matrix and $\sigma_{i},(i=x,y,z)$ are Pauli matrices. Then, the nonunitary Schr\"{o}dinger equation for effective NH BCS Hamiltonian [Eq. (5) in the main text] is rewritten in the following form:
\begin{align}
  \begin{bmatrix}
    H_{11} & H_{12} \\ H_{21} & H_{22}
  \end{bmatrix}\begin{bmatrix}
                 \Psi_{1,R} \\ \Psi_{2,R}
               \end{bmatrix} & = E \begin{bmatrix}
                                     \Psi_{1,R} \\ \Psi_{2,R}
                                   \end{bmatrix}. \label{NHhoneySupple_2by2Hamiltonian_rep_eq}
\end{align}
Here, $\Psi_{i, R}(i=1,2)$ constructs the right eigenvector of the Hamiltonian. From Eq.~\eqref{NHhoneySupple_2by2Hamiltonian_rep_eq}, one obtains
\begin{align}
  \begin{bmatrix}
    H_{11} + H_{12}(H_{11} + I)^{-1}H_{21}
  \end{bmatrix}\Psi_{1,R}  & = E\Psi_{1,R}, \label{NHhoneySupple_eigenequation_Psi1_eq} \\
  \begin{bmatrix}
    -H_{11} - H_{21}(H_{11} - I)^{-1}H_{12}
  \end{bmatrix}\Psi_{2,R} & = E\Psi_{2,R}. \label{NHhoneySupple_eigenequation_Psi2_eq}
\end{align}
With the particle-hole symmetry, Eq.~\eqref{NHhoneySupple_eigenequation_Psi1_eq} is identical to Eq.~\eqref{NHhoneySupple_eigenequation_Psi2_eq} by replacing $E$ with $-E$. Solving Eqs.~\eqref{NHhoneySupple_eigenequation_Psi1_eq} and \eqref{NHhoneySupple_eigenequation_Psi2_eq}, we obtain the eigenvalues $E_{\bm{k},\pm} = \sqrt{(|\epsilon_{\bm{k}}| \pm \mu)^{2} + \Delta\bar{\Delta}}$. Then, the effective BCS Hamiltonian reads
\begin{align}
  H_{\text{eff}}^{\text{BCS}} & = \sum_{\bm{k}}\left[E_{\bm{k},+}\bar{a}_{\bm{k},\uparrow}a_{\bm{k},\uparrow}+ E_{\bm{k},-}\bar{b}_{\bm{k},\uparrow}b_{\bm{k},\uparrow} + E_{\bm{k},+}\bar{a}_{-\bm{k},\downarrow}a_{-\bm{k},\downarrow} + E_{\bm{k},-}\bar{b}_{-\bm{k},\downarrow}b_{-\bm{k},\downarrow} \right]
  + \frac{N\Delta\bar{\Delta}}{U} - \sum_{\bm{k},\alpha}(E_{\bm{k},\alpha} + \mu),
\end{align}
where the quasiparticle operators $\bar{a}_{\bm{k},\sigma}a_{\bm{k},\sigma},\bar{b}_{\bm{k},\sigma}b_{\bm{k},\sigma}$ are given by the following expressions:
\begin{align}
  \bar{a}_{\bm{k},\uparrow} & = u_{\bm{k},+} d_{\bm{k},-,\uparrow}^{\dagger} - \bar{v}_{\bm{k},+} d_{-\bm{k},-,\downarrow}, \label{NHhoneySupple_quasiparticle_operator_bar_a_eq} \\
  a_{-\bm{k},\downarrow}    & = v_{\bm{k},+} d_{\bm{k},-,\uparrow}^{\dagger} + u_{\bm{k},+} d_{-\bm{k},-,\downarrow}, \label{NHhoneySupple_quasiparticle_operator_a_eq}           \\
  \bar{b}_{\bm{k},\uparrow} & = u_{\bm{k},-} d_{\bm{k},+,\uparrow}^{\dagger} - \bar{v}_{\bm{k},-} d_{-\bm{k},+,\downarrow}, \label{NHhoneySupple_quasiparticle_operator_bar_b_eq} \\
  b_{-\bm{k},\downarrow}    & = v_{\bm{k},-} d_{\bm{k},+,\uparrow}^{\dagger} + u_{\bm{k},-} d_{-\bm{k},+,\downarrow}. \label{NHhoneySupple_quasiparticle_operator_b_eq}
\end{align}
Here, the coefficients are given by
\begin{align}
  u_{\bm{k},+}       & = \sqrt{\frac{-\mu + E_{\bm{k},+} - |\epsilon_{\bm{k}}|}{2E_{\bm{k},+}}},                                                               \\
  v_{\bm{k},+}       & = -\frac{\bar{\Delta}}{-\mu + E_{\bm{k,+}} -|\epsilon_{\bm{k}}|}\sqrt{\frac{-\mu + E_{\bm{k},+} - |\epsilon_{\bm{k}}|}{2E_{\bm{k},+}}}, \\
  \bar{v}_{\bm{k},+} & =  -\frac{\Delta}{-\mu + E_{\bm{k,+}} -|\epsilon_{\bm{k}}|}\sqrt{\frac{-\mu + E_{\bm{k},+} - |\epsilon_{\bm{k}}|}{2E_{\bm{k},+}}},      \\
  u_{\bm{k},-}       & = \sqrt{\frac{-\mu + E_{\bm{k},-} + |\epsilon_{\bm{k}}|}{2E_{\bm{k},-}}},                                                               \\
  v_{\bm{k},-}       & = -\frac{\bar{\Delta}}{-\mu + E_{\bm{k,+}} +|\epsilon_{\bm{k}}|}\sqrt{\frac{-\mu + E_{\bm{k},-} + |\epsilon_{\bm{k}}|}{2E_{\bm{k},-}}}, \\
  \bar{v}_{\bm{k},-} & = -\frac{\Delta}{-\mu + E_{\bm{k,+}} +|\epsilon_{\bm{k}}|}\sqrt{\frac{-\mu + E_{\bm{k},-} + |\epsilon_{\bm{k}}|}{2E_{\bm{k},-}}},
\end{align}
where we have introduced the operators $d_{\bm{k},s,\sigma}$ in the main text. Here, the cooefficients satisfy $u_{\bm{k},\pm}^{2} + v_{\bm{k},\pm}\bar{v}_{\bm{k},\pm}=1$. We note that, due to the non-Hermiticity of the mean-field Hamiltonian, quasiparticle operators $a_{\bm{k},\sigma}$ and $\bar{a}_{\bm{k},\sigma}$ are not Hermitian conjugate to each other and satisfy an anticommutation relation $\{a_{\bm{k},\sigma},\bar{a}_{\bm{k}',\sigma'}\} = \delta_{\bm{k},\bm{k}'}\delta_{\sigma,\sigma'},\{b_{\bm{k},\sigma},\bar{b}_{\bm{k}',\sigma'}\} = \delta_{\bm{k},\bm{k}'}\delta_{\sigma,\sigma'}$.

Taking these relations into account, we can define the BCS state as the vacuum state for the quasiparticle operators \eqref{NHhoneySupple_quasiparticle_operator_bar_a_eq}-\eqref{NHhoneySupple_quasiparticle_operator_b_eq}. Then, we sholud require $a_{\bm{k},\sigma}|\text{BCS}\rangle_{R} = 0,b_{\bm{k},\sigma}|\text{BCS}\rangle_{R} = 0,\bar{a}_{\bm{k},\sigma}|\text{BCS}\rangle_{L} = 0,\bar{b}_{\bm{k},\sigma}|\text{BCS}\rangle_{L} = 0$, where $|\text{BCS}\rangle_{R}$ and $|\text{BCS}\rangle_{L}$ are right and left BCS states. As a result, we arrive at BCS states introduced in the main text [Eqs. (6) and (7)], which satisfy the normalization condition ${}_{L}\langle \text{BCS}|\text{BCS}\rangle_{R}=1$. By substituting BCS states to the definition of the order parameters $\Delta = -(2U/N)\sum_{\bm{k}}{}_{L}\langle c_{-\bm{k}s\downarrow}c_{\bm{k}s\uparrow}\rangle_{R}, \bar{\Delta} = -(2U/N)\sum_{\bm{k}}{}_{L}\langle c_{\bm{k}s\uparrow}^{\dagger}c_{-\bm{k}s\downarrow}^{\dagger}\rangle_{R}$, we have
\begin{align}
  \frac{N\Delta}{U}       & = \Delta \sum_{\bm{k}} (\frac{1}{2E_{\bm{k},+}} + \frac{1}{2E_{\bm{k},-}} ), \label{NHhoneySupple_delta_gap_Ek_representation_eq}          \\
  \frac{N\bar{\Delta}}{U} & = \bar{\Delta} \sum_{\bm{k}} (\frac{1}{2E_{\bm{k},+}} + \frac{1}{2E_{\bm{k},-}} ). \label{NHhoneySupple_bardelta_gap_Ek_representation_eq}
\end{align}
Finally, we obtain the NH gap equation given in Eq. (9) in the main text by dividing Eqs.~\eqref{NHhoneySupple_delta_gap_Ek_representation_eq} and \eqref{NHhoneySupple_bardelta_gap_Ek_representation_eq} by $\Delta,\bar{\Delta}$. In the same way, the mean particle number is calculated as
\begin{equation}
  M = {}_{L}\langle \text{BCS}|\hat{N}|\text{BCS}\rangle_{R} = 2\sum_{\bm{k}} \left[\frac{E_{\bm{k},+} + |\epsilon_{\bm{k}}| + \mu}{2E_{\bm{k},+}} + \frac{E_{\bm{k},-} - |\epsilon_{\bm{k}}| + \mu}{2E_{\bm{k},-}} \right]. \label{NHhoneySupple_MeanParticleNumber_eq}
\end{equation}
For $\mu=0$, the relations $E_{\bm{k},+}=E_{\bm{k},-}$ and $M=N$ follow.

Here, we remark an important relation between $\Delta$ and $\bar{\Delta}$. Because of the symmetry of the effective NH Hamiltonian [Eq. (1) in the main text], $H_{\text{eff}}^{\dagger} = H_{\text{eff}}^{\ast}$ under the matrix representation in terms of Fock states is satisfied, and the left and right eigenstates $|E\rangle_{L},|E\rangle_{R}$ obey the following relations
\begin{align}
   & H_{\text{eff}}|E\rangle_{R}    = E|E\rangle_{R} \Leftrightarrow H_{\text{eff}}^{\ast} (|E\rangle_{R})^{\ast} = E^{\ast}(|E\rangle_{R})^{\ast},                         \\
   & {}_{L}\langle E|H_{\text{eff}} = {}_{L}\langle E|E \Leftrightarrow H_{\text{eff}}^{\dagger}|E\rangle_{L} = H_{\text{eff}}^{\ast}|E\rangle_{L} = E^{\ast}|E\rangle_{L}.
\end{align}
By comparing the above two equations, we obtain the relation $(|E\rangle_{R})^{\ast} = |E\rangle_{L}$. Furthermore, as the effective NH Hamiltonian $H_{\text{eff}}$ has the U(1) symmetry, spontaneous U(1) symmetry breaking occurs and its state become degenerate when the superfluid is formed. Then, the BCS states [Eqs. (6) and (7) in the main text] are consistent with the following relation:
\begin{align}
   & \Delta       = \Delta_{0}e^{i\theta},  \\
   & \bar{\Delta} = \Delta_{0}e^{-i\theta},
\end{align}
where $\Delta_{0}\in \mathbb{C}$ and $\theta$ is the U(1) phase. Choosing the special gauge $\theta=0$, we obtain $\Delta=\bar{\Delta}=\Delta_{0}$. This gauge is consistent with the Hermitian limit $\bar{\Delta}= \Delta \in \mathbb{R}$.

Regarding the condensation energy of the superfluid state, it is defined as
\begin{equation}
  E_{\text{cond}} = {}_{L}\langle H_{\text{eff}}^{\text{BCS}}\rangle_{R} - {}_{L}\langle H_{\text{eff}}^{\text{BCS}}\left(\Delta=0,\bar{\Delta}=0\right) \rangle_{R} \label{NHhoneySupple_condensation_energy_rep_eq}
\end{equation}
where ${}_{L}\langle  H_{\text{eff}}^{\text{BCS}}\rangle_{R}$ and ${}_{L}\langle H_{\text{eff}}^{\text{BCS}}\left(\Delta=0,\bar{\Delta}=0\right) \rangle_{R}$ are the energy of the superfluid state and the normal state, respectively. These are written by
\begin{align}
   & {}_{L}\langle  H_{\text{eff}}^{\text{BCS}}\rangle_{R} =  \frac{N\Delta\bar{\Delta}}{U} - \sum_{\bm{k},\alpha} \left(E_{\bm{k},\alpha}+\mu\right),                    \\
   & {}_{L}\langle H_{\text{eff}}^{\text{BCS}}\left(\Delta=0,\bar{\Delta}=0\right) \rangle_{R} = -\sum_{\bm{k},\alpha}\left(||\epsilon_{\bm{k}}|+\alpha \mu| +\mu\right).
\end{align}
When $\text{Re} E_{\text{cond}}<0$ is satisfied, the superfluid state is stable.

\cred{
  Here, we comment on the phase diagram of our model away from the half-filling. Solving Eqs. (9) and \eqref{NHhoneySupple_MeanParticleNumber_eq} selfconsistently, we obtain a solution of $\Delta_{0},\mu\in \mathbb{C}$. The phase transition between the normal and DS states occurs, where ELs emerge in the reciprocal space. Thus, the cusp on the phase boundary and the NH phase transition should appear even away from the half-filling.}

To understand the property of the gap equation on a honeycomb lattice around the critical point $U_{c}$, where the energy dispersion is approximated as the linear one, we analyze the NH fermionic superfluids by assuming that DOS is given by $D(\epsilon) = (\epsilon/\pi v_{\text{F}}^{2})|\epsilon|$ with the fermi velocity $v_{\text{F}}$ (also see DOS on a honeycomb lattice illustrated in Fig. 2 (c) in the main text). Under this assumption, the NH gap equation reduces to
\begin{equation}
  \frac{N}{2U} = \frac{N}{2\pi v_{\text{F}}^{2}}\left(\sqrt{\Delta_{0}^{2} + 2\pi v_{\text{F}}^{2}} -\Delta_{0} \right).
\end{equation}
Here, we set a branch cut of $\sqrt{z}$ to $z\in (-\infty,0)$ and assume that $\text{Re}\Delta_{0}$ is positive without loss of generality. Then, the gap equation has only a trivial solution $\Delta_{0}=0$ for $|U| < U_{c}$ The critical interaction strength is introduced as $U_{c}=(\sqrt{\pi/2})v_{\text{F}}$. Non-trivial solution $\Delta_{0}\neq 0$ exists for $|U| > U_{c}$, and the superfluid gap is given by
\begin{equation}
  \Delta_{0}=\frac{U^{2}-U_{c}^{2}}{U}.
\end{equation}
Around $U=U_{c}$, both the real part of the gap and the imaginary part of that increase linearly as $\text{Re} \Delta_{0}\propto U_{1} - U_{c},\text{Im} \Delta_{0} \propto \gamma/2$. We then calculate the condensation energy Eq.~\eqref{NHhoneySupple_condensation_energy_rep_eq} for linear DOS. In this case, Eq.~\eqref{NHhoneySupple_condensation_energy_rep_eq} is calculated as
\begin{equation}
  \frac{E_{\text{cond}}}{N} = \frac{\Delta^{2}}{U} - \frac{1}{3 U_{c}^{2}} [(\Delta^{2} + 4U_{c}^{2})^{3/2} - (\Delta^{3} + 8 U_{c}^{3})].
\end{equation}

\begin{figure}[h]
  \centering
  \includegraphics[width = 12cm]{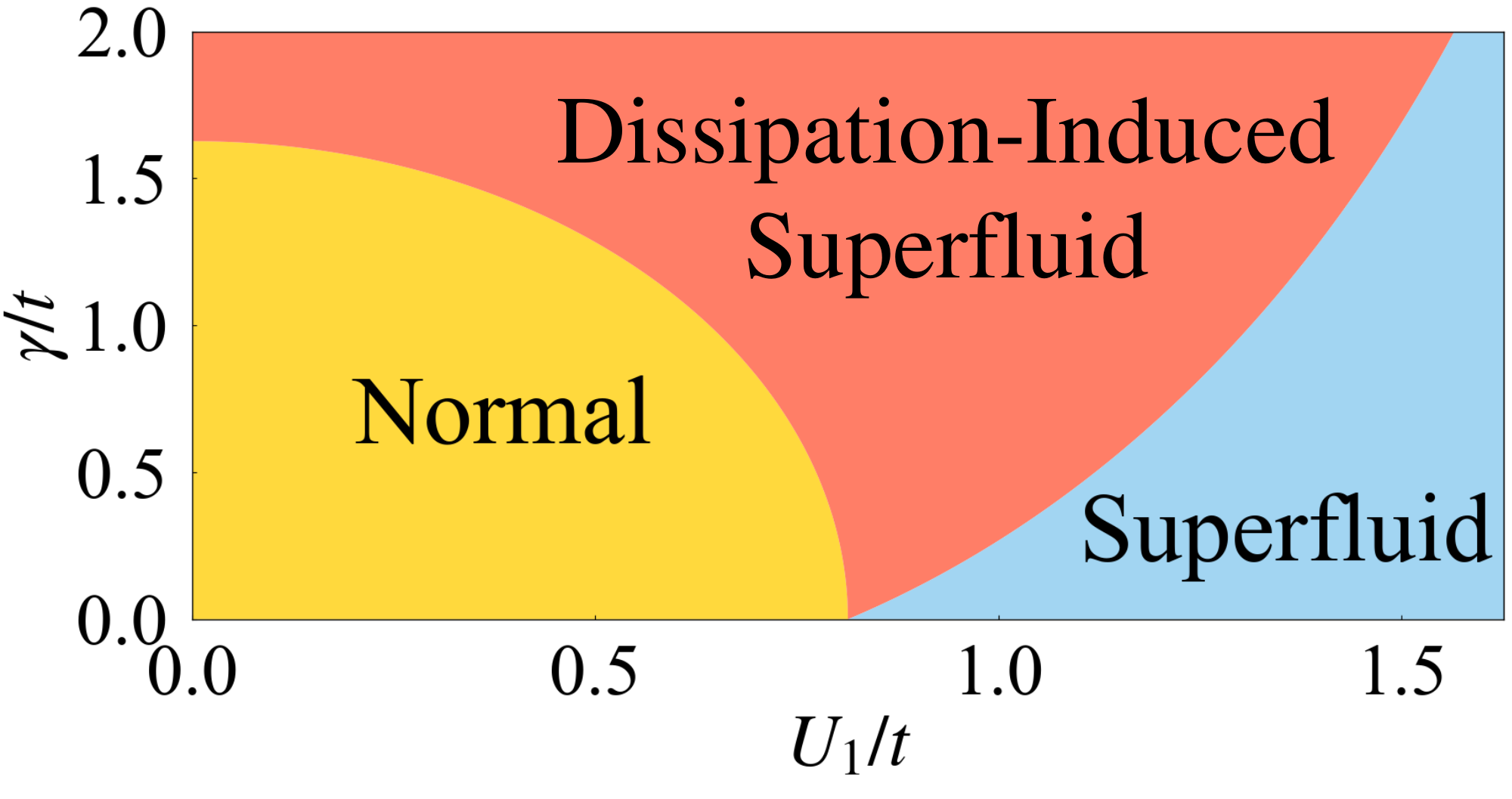}
  \caption{Phase diagram of the NH BCS model on a honeycomb lattice where DOS is approximated by $D(\epsilon)\propto \epsilon$. The red region shows the DS phase, the blue region corresponds to the superfluid phase, and the yellow region shows the normal phase. The phase transition from the normal phase to the DS phase in $U_{1}<U_{c}$ occurs, but the cusp on the phase boundary of the DS phase is not found.}
  \label{NHhoney_phasediagram_NHBCS_mu0_linearDOS_image}
\end{figure}

Finally, we show the phase diagram by the analytical calculation based on the linear DOS in Fig.~\ref{NHhoney_phasediagram_NHBCS_mu0_linearDOS_image}. For weak interaction where the normal state is stable in the Hermitian case, we find the non-trivial solution that gives the emergence of the superfluid. The phase boundary between the normal state and the DS state is given by $\gamma=2\sqrt{U_{c}^{2}-U_{1}^{2}}$. Moreover, around $U=U_{c}$, the DS state emerges for infinitesimal dissipation as already obtained by the numerical calculation shown in the main text.

Thus, this phase diagram roughly captures the feature that the phase transition between the normal state and the DS state occurs. However, the cusp on the phase boundary of the DS is not found. Such unconventional behavior of the phase boundary cannot be captured by the approximation with the use of the linear DOS and is caused by the interconnection between ELs and VHS.
\cred{It may be hard to obtain analytical formulas for the phase boundary near the cusp due to the logarithmic divergence of DOS.}

\section{\cred{The phase diagram of the NH BCS model on a square lattice}}
\begin{figure}[h]
  \centering
  \includegraphics[width=12cm]{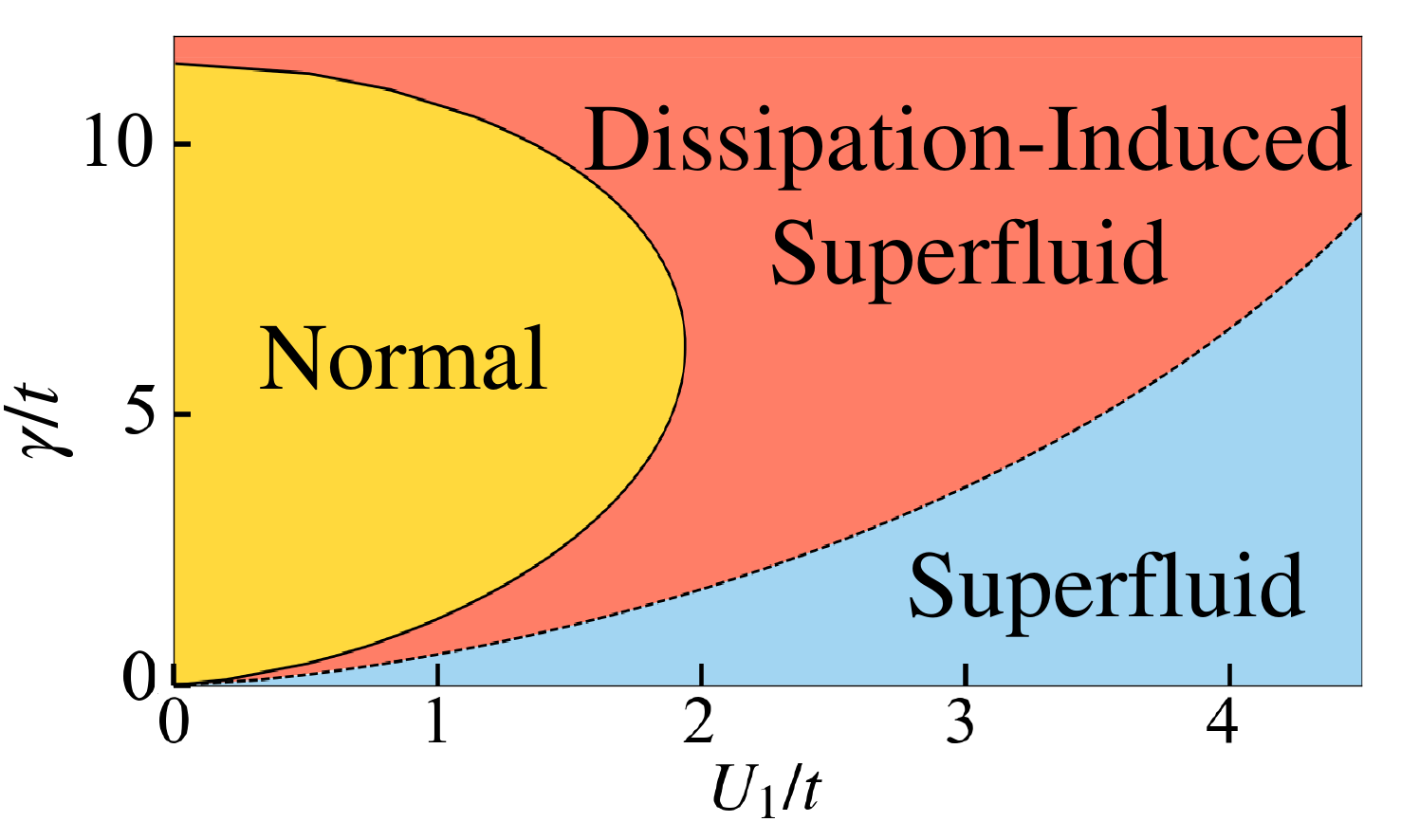}
  \caption{\cred{Phase diagram of the NH BCS model on a square lattice. The yellow region indicates the normal state, where only a trivial solution of the NH gap equation exists. The red region indicates to the DS state for which a non-trivial solution of the gap equation gives a positive real part of the condensation energy. The blue region indicates the superfluid phase corresponding to a non-trivial solution of the gap equation that gives a negative real part of the condensation energy.}}
  \label{NHhoneySupple_squarelattice_result_image}
\end{figure}

\cred{In this section, we show the phase diagram of the NH BCS model on a square lattice with $\mu=0$ in Fig.~\ref{NHhoneySupple_squarelattice_result_image}. We define the normal, DS and superfluid states in the same manner in our present paper. In the Hermitian limit $(\gamma=0)$, the critical attractive interaction $U_{c}$ which is defined by the phase transition point between the normal state and the superfluid state becomes zero on a square lattice. For weak attraction, the superfluid first breaks down with a small dissipation. In the case of strong dissipation, the nontrivial solution for DS state appears and ELs emerge at the phase boundary between the normal and DS states.
  For strong attraction, the phase transition between the DS and superfluid states appears. In the same manner as our present paper, both ELs and noninteracting DOS on a square lattice largely contribute to the NH gap equation. We find the absence of the cusp on the phase boundary because VHS is located at $\epsilon=0$ in the system on a square lattice.
  Therefore, our present result in Fig.1 in the main text is distinct from the result on a square lattice in Fig.~\ref{NHhoneySupple_squarelattice_result_image} because the position of VHS lies at a finite $\epsilon$ in the system on a honeycomb lattice. We expect that such a mechanism caused by the interplay between ELs and VHS is common to other systems with VHS at $\epsilon\neq 0$.}

\section{The behavior of the quasiparticle energy}
\begin{figure}[h]
  \centering
  \includegraphics[width = 12cm]{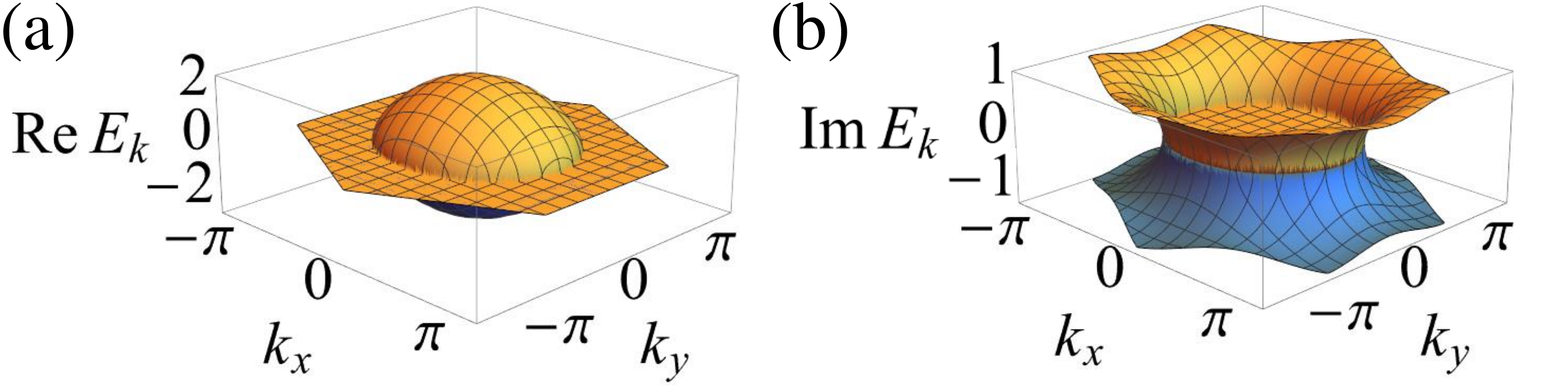}
  \caption{(a) Real part and (b) imaginary part of the quasiparticle energy spectrum $E_{\bm{k}}=\pm\sqrt{|\epsilon_{\bm{k}}|^{2} + \Delta_{0}^{2} }$. (Orange for the positive sign and blue for the negative sign). The parameter is set to $\Delta_{0}/t=1.8 i$. ELs form a circle at $E_{\bm{k}}=0$.}
  \label{NHhoney_quasiE_ReIm_image}
\end{figure}

In this section, we show the behavior of the real and the imaginary part of the quasiparticle energy $E_{\bm{k}}$ when ELs emerge. Figures \ref{NHhoney_quasiE_ReIm_image}(a) and (b) show the real part and the imaginary part of the quasiparticle energy $E_{\bm{k}}$ for $\Delta_{0}/t = 1.8i$, respectively. We see that ELs form a circle when both $\text{Re} E_{\bm{k}}$ and $\text{Im} E_{\bm{k}}$ become zero. Moreover, when ELs appear, we find that $\text{Im} E_{\bm{k}}$ takes a finite value only outside ELs. As the imaginary part of the quasiparticle energy corresponds to its lifetime, such a quasiparticle distribution implies that the lifetime is amplified for $\text{Im} E_{\bm{k}}>0$ in the particular region in momentum space. This characteristic behavior can be utilized to experimentally observe ELs \cite{yamamoto19}.

\section{The detail of the numerical calculation of the NH gap equation}
\begin{figure}[h]
  \centering
  \includegraphics[width = 12cm]{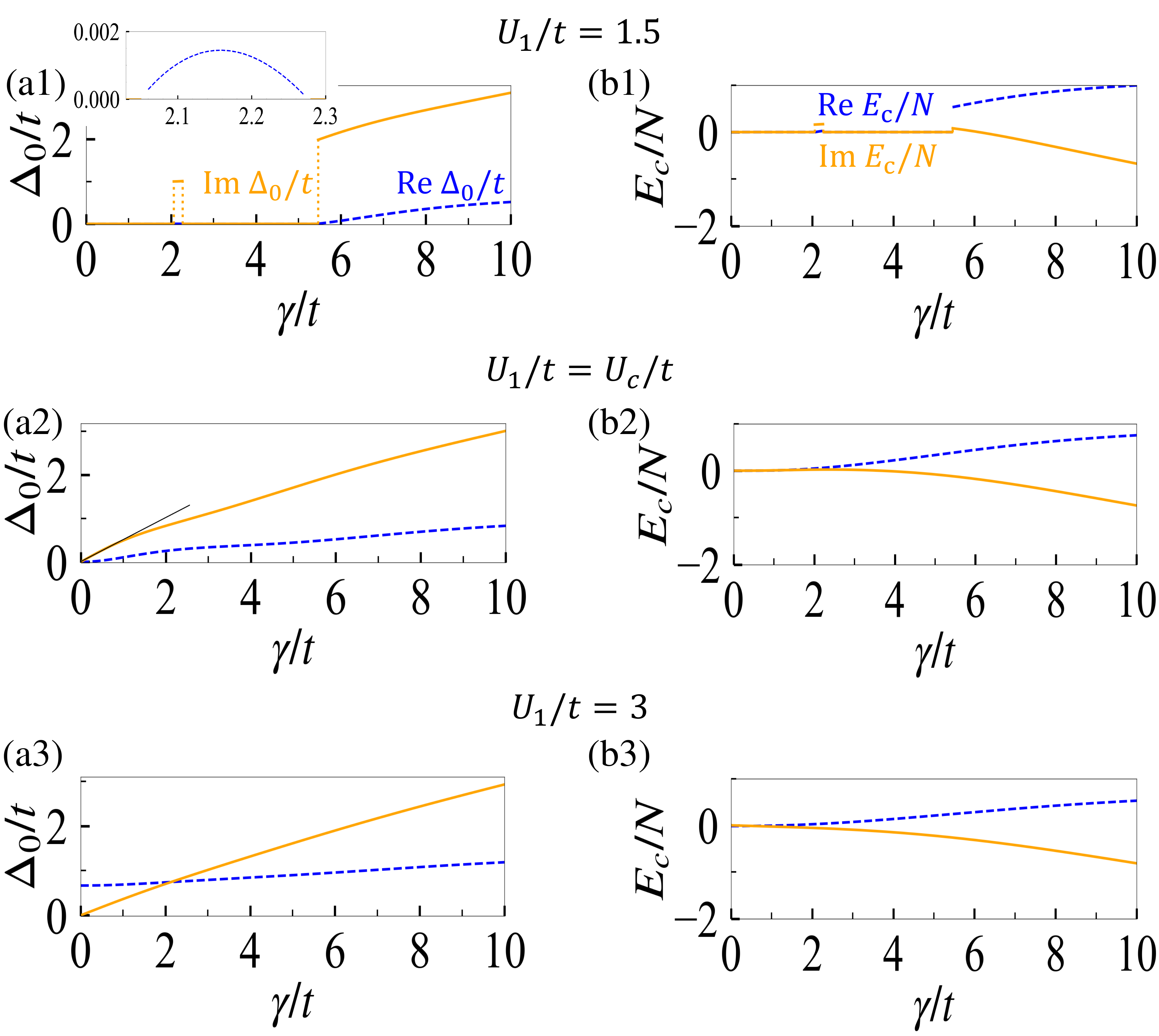}
  \caption{(a1)-(a3) Dashed (solid) lines represent the real (imaginary) part of
  the superfluid gap and (b1)-(b3) condensation energy for $U_{1}/t=1.5,2.23,3.0$, respectively. The imaginary part of the superfluid gap first increases linearly at $U_{1}=U_{c}$ as we introduce dissipation [see black line in (a2)]}
  \label{NHhoneySupple_D_Ec_behaviour_image}
\end{figure}

In this section, we show the detailed results obtained by the numerical calculations for the superfluid gap [Eq. (9) in the main text] and the condensation energy. We show the results for weak interactions $U_{1}/t = 1.5$ as illustrated in Figs.~\ref{NHhoneySupple_D_Ec_behaviour_image}(a1) and (b1). As $\gamma$ increases, the real part and the imaginary part of the superfluid gap appear and instantly disappear at $\gamma/t\sim 2$. This anomalous behavior is seen around the cusp on the phase boundary between the normal state and the DS state, which is caused by the interplay between ELs and VHS. We find the robustness of the superfluid state for strong dissipation, which indicates that QZE occurs. This superfluid phase for strong dissipation corresponds to the metastable superfluid because $\text{Re}E_{\text{cond}}>0$ is satisfied. For the critical interaction strength $U=U_{c}$ shown in Figs.~\ref{NHhoneySupple_D_Ec_behaviour_image}(a2) and (b2), the phase transition between the normal state and the DS state occurs for infinitesimal dissipation rate. As shown by the black line in Fig. 3, the imaginary part of the superfluid gap is proportional to $\gamma$. Here, we recall that the linear dependence of the superfluid gap with respect to the interaction strength in the Hermitian system indicates that the physics is determined by that governed by the Dirac dispersion relations. Therefore, we see that the phenomena around the critical interaction strength $U=U_{c}$ is governed by those described by the Dirac dispersion relations even in the NH system. On the other hand, for strong interaction $U_{1}/t=3.0$ illustrated in Fig.~\ref{NHhoneySupple_D_Ec_behaviour_image}(a3) and (b3), the superfluid gap appears for any dissipation strength. This result indicates that the cooper pair can survive for strong dissipation, which is reminiscent of the behavior of the cubic lattice \cite{yamamoto19}. However, the real part of the condensation energy is only negative for small dissipation, which still leads to the energetical stability of the superfluid.

We note that the imaginary part of the condensation energy seems to be positive around the cusp on the phase boundary of the DS state [see Fig.~\ref{NHhoneySupple_D_Ec_behaviour_image}(b1)]. Because of the completely positive and trace preserving map that describes the Lindblad master equation, the imaginary part of the expectation value of the effective Hamiltonian derived from the Lindblad equation should be negative. Then, this indicates that the imaginary part of the condensation energy may also be negative. Thus, we predict that the positivity of the imaginary part of the condensation energy may be an artifact of the mean-field approximation. We expect that the negativity of the imaginary part of the condensation energy can be recovered if we appropriately include the Hartree term ignored in the BCS approximation.


Finally, we remark on the method of the self-consistent calculation used to solve the NH gap equation [Eq. (9) in the main text]. In general, we have difficulty with the convergence of the numerical calculation due to strong non-Hermiticity, particularly when we approach ELs. By analyzing the convergence behavior of the superfluid gap by solving the NH gap equation, we find that the succession of the superfluid gap forms the elliptic shape when it approaches a correct solution. Then, by choosing the center of the ellipse as the initial state of the numerical calculation and using the similar technique used in the method of successive over-relaxation \cite{hadjidimos00sor}, we successfully obtain the convergent solution.

\section{\cred{Experimental protocols for our systems}}
\cred{Here, we explain an experimental setup for testing our system. The dynamics of the effective NH Hamiltonian in our model can be realized in ultracold atoms. We consider a quasi-stationary state which is expected to be realized after a finite-time relaxation process under the NH dynamics. This requires that the timescale of the relaxation should be shorter or comparable to $1/\gamma$. Then quantum-jump process is dominated by the loss rate $\gamma$. For a BCS superfluid, the relaxation towards a quasi-equilibrium state, which is governed by hopping of atoms to neighboring sites, proceeds in $\gamma \simeq t$. Under this condition, the NH dynamics can be observed. The particle number can be measured by using a quantum-gas microscope~\cite{ashida16} thanks to the recent experimetal progress in ultracold atoms and the non-Hermitian dynamics is faithfully realized.}

\begin{figure}[h]
  \centering
  \includegraphics[width=12cm]{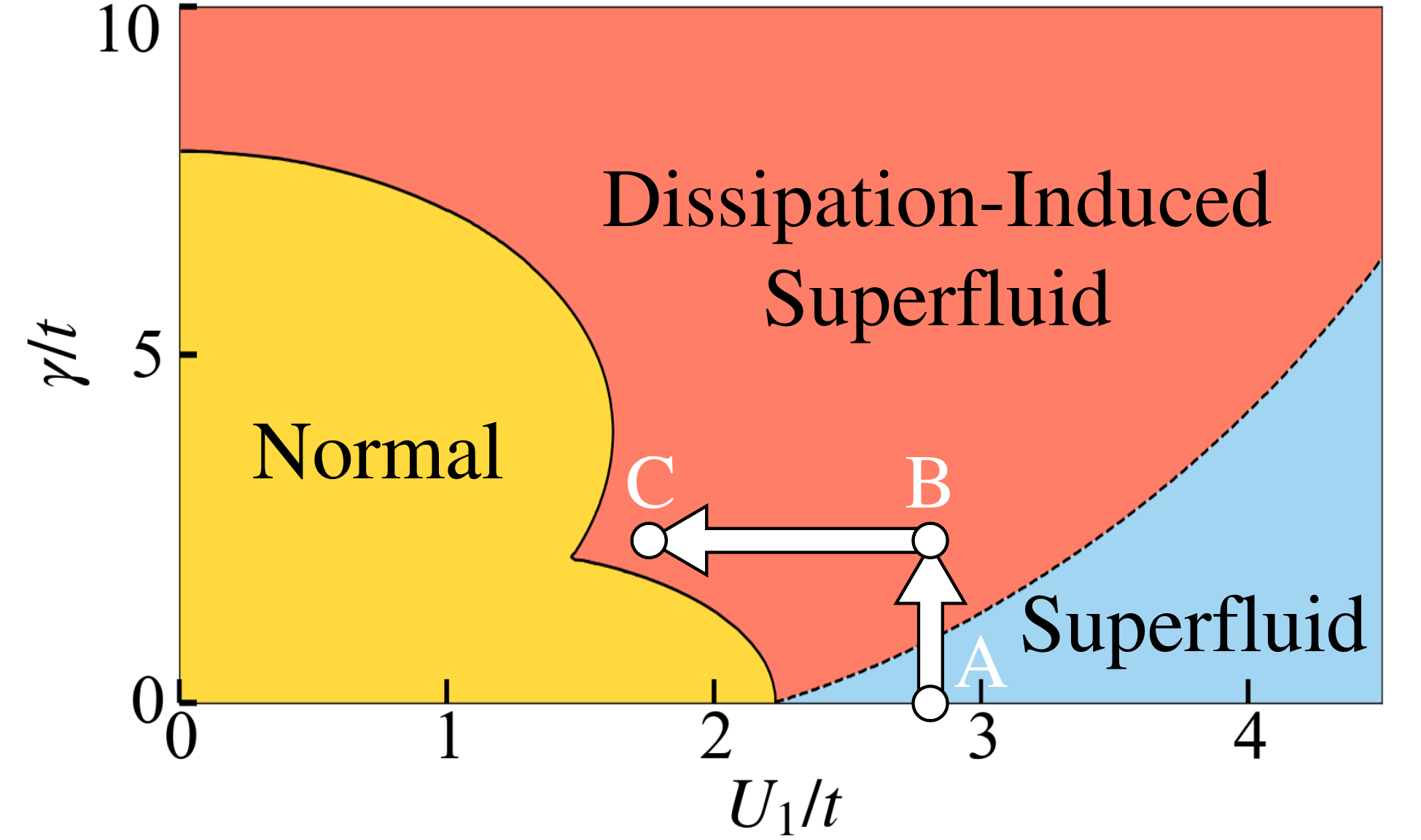}
  \caption{\cred{Experimental protocol for the detection of the phase boundary between the DS state and the normal state. We first prepare a superfluid state for large attraction $U_{1}$ (A) and then introduce large dissipation $\gamma$ by using a photoassociation technique (B). Finally, the system will reach the phase boundary by ramping down $U_{1}$ (C).}}
  \label{NHhoneyreport_protocols_DS_and_normal_image}
\end{figure}

\cred{We explain a detailed experimental protocol for our system. The experimental protocol for the detection of the phase boundary between the DS state and the normal state is illustrated in Fig.~\ref{NHhoneyreport_protocols_DS_and_normal_image}. We first prepare a superfluid state for large attraction $U_{1}$ [(A) in Fig.~\ref{NHhoneyreport_protocols_DS_and_normal_image}]. Then, by using photoassociation techniques~\cite{tomita17}, we introduce large dissipation $\gamma/t\sim 2$ [(B) in Fig.~\ref{NHhoneyreport_protocols_DS_and_normal_image}]. Lastly, we decrease the attraction $U_{1}$ by means of a Feshbach resonance [(C) in Fig.~\ref{NHhoneyreport_protocols_DS_and_normal_image}]. The relaxation timescale is estimated by the inverse hopping rate as $1/t$. However, the tunneling of atoms is suppressed by the QZE under a large dissipation and the dissociation of molecules after decreasing $U_{1}$ is delayed. Consequently, atoms in the system will survive even in the presence of dissipation. Furthermore, decreasing $U_{1}$, the order parameter suddenly vanishes, and we can detect the phase boundary between the normal and DS states.}

\end{document}